\documentclass[aps,twocolumn]{revtex4}
\usepackage{graphicx}    \usepackage{rotating}    \usepackage{dcolumn}
\usepackage{epsfig} 
\usepackage{amssymb}        
\usepackage{amsmath}     
\begin{document}

\title{The staircase structure of the Southern Brazilian Continental
    Shelf}
\author{M. S. Baptista}
\affiliation{Centro de Matem\'atica da Universidade do Porto, Rua do Campo
Alegre 687, 4169-007 Porto, Portugal}
\author{L. A. Conti} \affiliation{Escola de Artes Ci\^encias e Humanidades,
Universidade de S\~ao Paulo, Av. Arlindo Bettio 1000, 03828-000, S\~ao Paulo,
Brasil}

\begin{abstract}
We show some evidences that the Southeastern Brazilian Continental
Shelf (SBCS) has a devil's staircase structure, with a sequence of
scarps and terraces with widths that obey fractal formation rules.
Since the formation of these features are linked with the sea level
variations, we say that the sea level changes in an organized
pulsating way.  Although the proposed approach was applied in a
particular region of the Earth, it is suitable to be applied in an
integrated way to other Shelves around the world, since the analyzes
favor the revelation of the global sea level variations.
\end{abstract}
\maketitle
\section{Introduction}

During the late quaternary period, after the last glacial maximum
(LGM), from 18 kyears ago till present, a global warming was
responsible for the melting of the glaciers leading to a fast increase
in the sea level. In approximately 13 kyears, the sea level rised up
to about 120 meters, reaching the actual level.  However, the sea
level did not go up in a continuous fashion, but rather, it has
evolved in a pulsatile way, leaving behind a signature of what
actually happened, the Continental Shelf, i.e.  the seafloor.

Continental shelves are located at the boundary with the land so that
they are shaped by both marine and terrestrial processes. Sea-level
oscillations incessantly transform terrestrial areas in marine
environments and vice-versa, thus increasing the landscape complexity
\cite{lambeck:2001}. 
The presence of regions with abnormal slope as well as the presence of
terraces on a Continental Shelf are indicators of sea level positions
after the Last Glacial Maximum (LGM), when large ice sheets covered
high latitudes of Europe and North America, and sea levels stood about
120-130m lower than today
\cite{heinrich:1988}. Geomorphic processes responsible for the formation of
these terraces and discontinuities on the bottom of the sea topography
are linked to the coastal dynamics during eustatic processes
associated with both erosional or depositional forcing (wave cut and
wave built terraces respectively
\cite{goslar:2000}).

The irregular distribution of such terraces and shoreface sediments is
mainly controlled by the relationship between shelf paleo-physiography
and changes on the sea level and sediment supply which reflect both
global and local processes. Several works have dealt with mapping and
modeling the distribution of shelf terraces in order to understand the
environmental consequences of climate change and sea level variations
after the LGM
\cite{adams:1999,andrews:1998,broecker:1998}.

In this period of time the sea-level transgression was punctuated by
at least six relatively short flooding events that collectively
accounted for more than 90m of the 120m rise. Most (but not all) of
the floodings appear to correspond with paleoclimatic events recorded
in Greenland and Antarctic ice-cores, indicative of the close coupling
between rapid climate change, glacial melt, and corresponding
sea-level rise \cite{taylor:1997}.

In this work, we analyze data from the Southeastern Brazilian
Continental Shelf (SBCS) located in a typical sandy passive margin
with the predominance of palimpsests sediments. The mean length is
approximately 250km and the shelfbreak is located at 150m depth. It is
a portion of a greater geomorphologic region of the southeastern
Brazilian coast called S\~ao Paulo Bight, an arc-shaped part of the
southeastern Brazilian margin. The geology and topography of the
immersed area are very peculiar, represented by the Mesozoic/Cenozoic
tectonic processes that generated the mountainous landscapes known as
"Serra do Mar". These landscapes (with mean altitudes of 800m) have a
complex pattern that characterize the coastal morphology, and leads to
several scarps intercalated with small coastal plains and pocket
beaches.

This particular characteristic determines the development of several
small size fluvial basins and absence of major rivers conditioning low
sediment input, what tends to preserve topographic signatures of the
sea-level variations.

For the purpose of the present study, we select three parallel
profiles acquired from echo-sounding surveys, since for all the
considered profiles, the same similar series of sequences of terraces
were found.  These profiles
\cite{furtado:1992,conti:2001,correa:1996} are transversal to the coastline
and the isobaths trend, and they extend from a 20m to a 120m depth.

The importance of understanding the formation of these ridges is that
it can tell us about the coastal morphodynamic conditions, inner shelf
processes and about the characteristics of periods of the sea level
regimes standstills (paleoshores). In particular, the widths of the
terraces are related to the time the sea level "stabilized". All this
information is vital for the better understanding of the late
quaternary climate changes dynamic.

We find relations between the widths of the terraces that follow a
self-affine pattern description.  These relations are given by a
mathematical model, which describes an order of appearance for the
terraces. Our results suggest that this geomorphological structure for
the terraces can be described by a devil's staircase
\cite{mandelbrot:1977}, a staircase with infinitely many steps in
between two steps. This property gives the name "devil" to the
staircase, once an idealized being would take an infinite time to go
from one step to another. So, the seafloor morphology is self-affine
(fractal structure) as reported in Ref. \cite{herzfeld,goff}, but
according to our findings, it has a special kind of self-affine
structure, the devil's staircase structure.

A devil's staircase as well as other self-affine structure are the
response of an oscillatory system when excited by some external
force. The presence of a step means that while varying some internal
parameter, the system preserves some averaged regular behavior, a
consequence of the stable frequency-locking regime between a natural
frequency of the system and the frequency of the excitation.  This
staircase as well as other self-affine structures are characterized by
the presence of steps whose widths are directly related to the
rational ratio between the natural frequency of the system and the
frequency of the excitation.

In a similar fashion, we associate the widths of the terraces with
rational numbers that represent two hypothetical frequencies of
oscillation which are assumed to exist in the system that creates the
structure of the SBCS, here regarded as the sea level dynamics (SLD),
also known as the sea level variations. Then, once these rational
numbers are found, we show that the relative distances between triples
of terraces (associated with some hypothetical frequencies) follow
similar scalings found in the relative distance between triples of
plateaus (associated with these same frequencies) observed in the
devil's staircase.

The seafloor true structure, apart from the dynamics that originated
it, is also a very relevant issue, specially for practical
applications. For example, one can measure the seafloor with one
resolution and then reconstruct the rest based on some modeling
\cite{mareschal}. As we show in this work (Sec. \ref{model}), a
devil's staircase structure fits remarkably well the experimental
data.

Our paper is organized as follows. In Sec. \ref{data}, we describe the data to
be analyzed. In Sec. \ref{devil}, we describe which kind of dynamical systems
can create a devil's staircase and how one can detect its presence in
experimental data based on only a few observations. In Sec.
\ref{devil_in_data}, we show the evidences that led us
to characterize the SBCS as a devil's staircase, and in
Sec. \ref{model} we show how to construct seafloor profiles based on
the devil's staircase geometry. Finally, in Sec.
\ref{conclusao}, we present our conclusions, discussing also possible
scenarios for the future of the sea level dynamics under the
perspective of our findings.

\section{Data}\label{data}

\begin{figure}
  \centerline{\hbox{\psfig{file=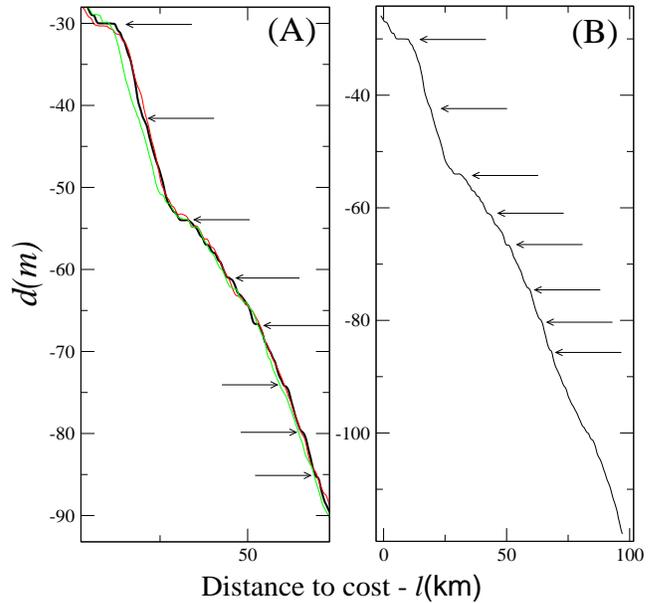,height=9.0cm,width=9cm}}}
  \caption{[Color online] Profiles (depth versus the distance to the cost) of the
  Southeastern Brazilian Continental Shelf. The arrows indicate the terraces
  considered in our analyzes. The profile shown with a thick black line is the
  profile chosen for our derivations, reproduced also in (B). The other two
  profiles had their original position of the two axes shifted by a constant
  value such that one can also identify the terraces observed in the chosen
  profile in these other two. Note that a translation of the profiles by a
  constant value has no effect on any of the scalings observed. The reason of
  this mismatch between the profiles is due to the local geometry of the
  cost at the time the sea reached that level.}
\label{meco_fig1}
\end{figure}

The data consists of the tree profiles given in Fig.
\ref{meco_fig1}(A-B). The profile considered for our analyzes 
is shown in Fig.  \ref{meco_fig1}(B), where we show the Continental Shelf of
the State of S{\~a}o Paulo, in a transversal cut in the direction: inner shelf
("cost")
$\rightarrow$ shelfbreak ("open sea").  The horizontal axis represents
the distance to the cost and the vertical axis, the sea level (depth), $d$. We
are interested in the terraces widths and their respective depths.

The profiles shown in Fig. \ref{meco_fig1} were the result of a smoothing
(filtering) process from the original data collected by Sonar
\cite{note3}. The smoothing process is needed to eliminate from the measured
data the influence of the oscillations of the ship where the sonar is located
and local oscillations on the sea floor probably due to the stream flows.

\begin{figure}
\centerline{\hbox{\psfig{file=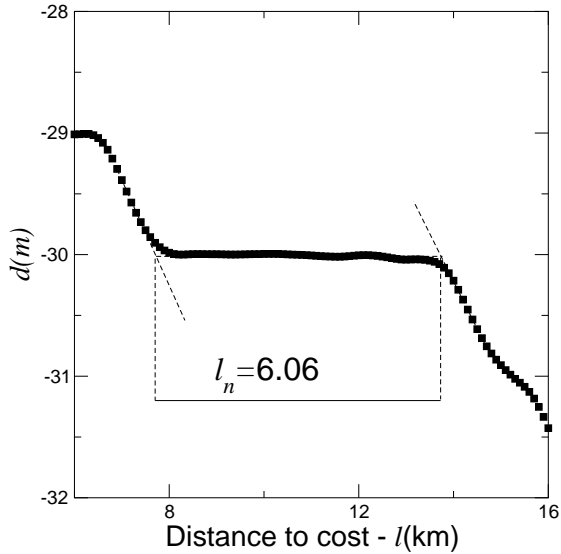,width=9.0cm}}} 
\caption{Reconstruction of the terraces.  The width of the terrace is
  given by the minimal distance between the two parallel dashed lines. }
\label{meco_fig2}
\end{figure}

Smaller topographic terraces could be smoothed or masked due to several
processes such as: coastal dynamic erosional during sea-level rising, Holocene
sediment cover, erosional processes associated with modern hidrodynamic
pattern (geostrophic currents). For that reason we only consider the largest
ones, as the ones shown in Fig. \ref{meco_fig2}, (located at $d=-30.01m$ with
the width of $l=6.06km$).  As one can see, the edges of the terraces are not
so sharp as one would expect from a staircase plateau.  Again, this is due to
the action of the sea waves and stream flows throughout the time. To
reconstruct what we believe to be the original terrace, we consider that its
depth is given by the depth of the middle point, and its width is given by the
minimal distance between two parallel lines placed along the scarps of the
terrace edges. Using this procedure, we construct Table \ref{table1} with the
largest and more relevant terraces found.

We identify a certain terrace introducing a lower index $n$ in $l$ and $d$,
according to their chronological order of appearance. More recent appearance
(closer to the cost, less deep) smaller is the index $n$.  We consider the
more recent data to have a zero distance from the cost, but in fact, this data
is positioned at about 15km away from the shore, where the bottom of the sea
is not affected by the turbulent zone caused by the break out of the
waves. The profile of Fig.  \ref{meco_fig1}(B) was the one chosen among the
other tree profiles because from it we could more clearly identify the
largest number of relevant terraces
\cite{note3}.

\begin{center}
\begin{table}
\begin{tabular}{c|c|c} 
  $n$ & $d_n (m)$  & $l_n (km)$   \\ \hline
  1 & -30.01 & 6.06 $\pm$ 0.05  \\
  2 & -41.86 & 1.59  $\pm$ 0.05 \\
  3 & -54.01 & 2.93  $\pm$ 0.05 \\ 
  4 & -61.14 & 1.73 $\pm$ 0.05 \\
  5 & -66.69 & 2.21 $\pm$ 0.05 \\ 
  6 & -74.33 & 0.80 $\pm$ 0.1 \\
  7 & -79.75 & 0.80 $\pm$ 0.1 \\
  8 & -85.30 & 0.80 $\pm$ 0.1 \\ \hline
\end{tabular}
\caption{Terrace widths and depths. While the depths present 
  no representative deviation, the deviation in the 
  widths become larger for deeper terraces. The deviation
  in the widths is estimated by calculating the widths
  assuming many possible configurations 
  between the placement of the two parallel lines
  used to calculate the widths.}
\label{table1}
\end{table}
\end{center}

\section{The Devil's Staircase}\label{devil}

Frequency-locking is a resonant response occurring in systems of
coupled oscillators or oscillators coupled to external forces. The
first relevant report about this phenomenon was given by Christian
Huygens in the 17$^th$ century. He observed that two clocks back to
back in the wall, set initially with slightly different frequencies,
would have their oscillations coupled by the energy transfer
throughout the wall, and then they would eventually have their
frequencies synchronized.  Usually, we expect that an
harmonic, $P_{w1}$, of one oscillatory system locks with an harmonic,
$Q_{w2}$, of the other oscillatory system, leading to a locked system
working in the rational ratio $P/Q$ \cite{jensen:1984}.

To understand what is the dynamics responsible for the onset of a
frequency-locked oscillation, that is, the reasons for which a system
either locks or unlocks, we present the simplest model one can come up
with to describe a more general oscillator. This model is described by
an angle $\theta$, which is changed (after one period) to the angle
$f(\theta)$. So, $f(\theta)=\theta + \Omega$. In order to introduce an 
external force in the oscillator modeling also possible physical interactions
with other oscillators, a resonant term, $g$, is added into this
model, resulting in the following model

\begin{equation}
f(\theta)=\theta+\Omega-g(\theta,K)  \mbox{\ \ \ } (mod \mbox{\ } 1), 
\label{circle_map}
\end{equation}
\noindent
where 

\begin{equation}
g(\theta,K) = \frac{K}{2\pi}\sin{2 \pi \theta}.
\label{g_theta}
\end{equation}

Despite this map simplicity, the same can not be said about its
complexity \cite{argyris:1994}.  Arnold (see ref. \cite{arnold:1965})
studied this map in detail aiming to understand how an oscillatory
system would undergo into periodic stable state when perturbed by an
external perturbation.

For $K=0$, Eq. (\ref{circle_map}) represents a pure rotation, which is
topological equivalent to a twice continuously differentiable, orienting
preserving mapping of the circle onto itself [Theorem of Denjoy, see
Ref. \cite{arnold:1988}].  Therefore, the simple Eq. (\ref{circle_map}) can be
considered as a model for many types of oscillatory systems. In fact, Eq.
(\ref{circle_map}) represents a more complicated system, a three-dimensional
torus with frequencies $w_1$ and $w_2$, when viewed by a Poincar{\'e}
map. Thus, $\Omega$ in Eq. (\ref{circle_map}) represents the ratio $w_1/w_2$.
When $w_1/w_2=p/q$ (with $p \leq q$) is rational, this map has a period $p$
motion and its trajectory, i.e.  the value of $\theta$, assumes the same value
after $q$ iterations. For $K=0$, the so called winding number $W$ is exactly equal to
$\Omega$, i.e. $W$=$p/q$.

For $K \neq 0$ (non-linear case) $W$ is defined by
{\small 
\begin{equation}
  W(\Omega,K)= \lim_{n \to \infty} \frac{h(\theta_0,K)+h(\theta_1,K)+\ldots+h(\theta_{n-1},K)}{n},
\label{winding_number}
\end{equation}}
\noindent
where
\begin{equation}
h(\theta,K)=\Omega+g(\theta,K).
\label{h} 
\end{equation}

For $K<1$, Eq. (\ref{circle_map}) is monotonic and invertible.  For
$K=1$, it develops a cubic inflection point at $\theta=0$. The map
is still invertible but the inverse has a singularity. For $K>1$ the
map is non-invertible.

\begin{figure}
\centerline{\hbox{\psfig{file=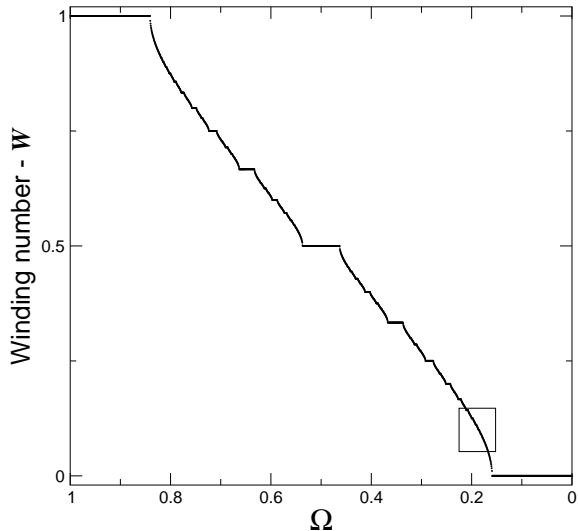,width=8cm}}} 
\caption{A complete devil's staircase, obtained from Eq.
  (\ref{circle_map}), for $K$=1.}
\label{meco_fig6}
\end{figure}

Arnold wanted to understand how periodic oscillations would appear as
one increases $K$ from zero to positive values.  He observed that a
quasi-periodic oscillation, for an irrational $\Omega$ and $K=0$,
would turn into a periodic oscillation as one varies $K$, from zero
to positive values. He demonstrated that a periodic oscillation has
probability zero of being found for $K=0$ (rational numbers set is
countable while the irrational numbers set is uncountable) and
positive probability of being found for $K>0$. He also observed that
fixing a positive value $K$, the winding number $W$ [Eq.
(\ref{winding_number})] is a continuous but not differentiable
function of $\Omega$, as one can see in Fig. \ref{meco_fig6}, forming
a stair-like structure.

If $W(\Omega,K)$ is rational, it can be represented by the ratio between two
integer numbers as $W=\frac{P}{Q}$. At this point, the frequency $\Omega$ and
the frequency of the function $g$ are locked, producing the phenomenon of
frequency-locking, when $W(K,\Omega)$ does not change its value within an
interval $\Delta \Omega$ (a plateau) of values of $\Omega$. In fact, smaller
is the denominator $Q$, larger is the interval $\Delta
\Omega$. 

As one changes $\Omega$, plateaus for $W$ rational appear following a natural
order described by the Farey mediant. Given two plateaus that represent a
$\frac{P_1}{Q_1}$ and a $\frac{P_3}{Q_3}$ winding numbers, with plateau widths
of $\Delta \Omega_1$ and $\Delta \Omega_3$, respectively, there exists another
plateau positioned at a winding number $W$ within the interval
$[\frac{P_1}{Q_1},\frac{P_3}{Q_3}]$ given by

\begin{equation}
\frac{P_2}{Q_2}=\frac{P_1+P_3}{Q_1+Q_3}.
\label{farey_mediant}
\end{equation}
\noindent
The Farey mediant gives the rational with the smallest integer denominator
that is within the interval $[\frac{P_1}{Q_1},\frac{P_3}{Q_3}]$.  Therefore,
the $\Delta \Omega (\frac{P_2}{Q_2})$ plateau is smaller than $\Delta \Omega
(\frac{P_1}{Q_1})$ and $\Delta \Omega (\frac{P_3}{Q_3})$, but is bigger than
any other possible plateau.  Organizing the rationals according to the Farey
mediant creates an hierarchical level of rationals, which are called Farey
Tree. The plateaus $\Delta \Omega_1$ and $\Delta \Omega_3$ are regarded as the
parents, and $\Delta \Omega_2$ as the daughter plateau.

The interesting case of Eq. (\ref{circle_map}) for our purpose, is exactly
when $K$=1.  For that case, one can find periodic orbits with any possible
rational winding number, as one varies $\Omega$.  What results in a zero
(probability) measure of finding quasi-periodic oscillation in Eq.
(\ref{circle_map}), by a random chosen of the $\Omega$ value.  Also, for
$K>1$, due to the overlap of resonances (periodic oscillation), chaos is
possible.

The devil's staircase can be fully characterized by the relations between the
plateaus widths, and the relations between the gaps between two of them. While
the plateau widths are linked to the probability one has to find periodic
oscillations, the gaps widths between plateaus are linked to the probability one has to
find quasi-periodic oscillations, in Eq. (\ref{circle_map}). 

There are many scaling
laws relating the plateau widths
\cite{jensen:1983,jensen:1984,cvitanovic:1985}.  There are local
scalings, which relate the widths of plateaus that appear close to a specific
winding number, for example the famous golden mean
$W_G=\frac{\sqrt{5}-1}{2}$. However, we will focus our attention in the global
scalings, which can be experimentally observed, and only for the case where
$K$=1.  For this case, we are interested in two scalings. The one that relates
the plateau widths with the respective winding numbers in the form
$\frac{1}{Q}$ (the largest plateaus), and the one that describes the structure
of the complementary set to the plateaus $\Delta \Omega$, i.e. the structure
of the gaps between plateaus. The structure of the plateaus is a Cantor
set as well as the structure of the complementary set. Therefore, a
characterization of these sets can be done in terms of the fractal dimension
$D_0$
\cite{farmer:1983} of the complementary set.

The first scaling is 
\cite{jensen:1984}

\begin{equation}
\Delta \Omega(\frac{1}{Q}) \propto \frac{1}{Q^\gamma} \mbox{\ \ } (\gamma>3),
\label{scaling_1}
\end{equation}
\noindent
The second scaling relates the widths of the complementary set as one goes to
smaller and smaller scales. These widths are related to a power-scaling law
whose coefficient $D_0$ is the fractal dimension of the complementary set.
For $K=1$, we have that the fractal dimension of the complementary set is $D_0
\cong 0.87$. 
This is an universal scaling.  Since the complementary set of the plateaus
represents the irrational rotations, the smaller is its fractal dimension, the
smaller is the probability of finding quasi-periodic oscillation.

\begin{figure}
\centerline{\hbox{\psfig{file=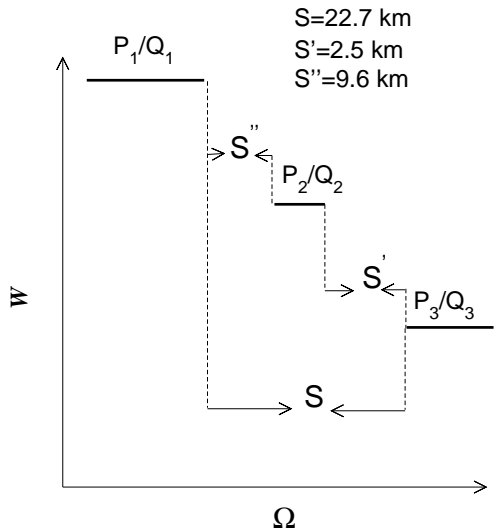,width=8.0cm}}} 
\caption{Representation of the gaps $S^{\prime\prime}$, 
$S^{\prime}$, and $S$, used to estimate the fractal 
dimension $D_0$ of the complementary set, using 
Eq. (\ref{scaling_2}).}
\label{meco_fig6_1}
\end{figure}

For experimental data, the determination of $D_0$ is difficult to
obtain because the dimension measures a microscopic quantity of the
plateaus widths, and in experimental data one can only observe the
largest plateaus.  Fortunately, an approximation $D^{\prime}$ for
$D_0$ can be obtained from the largest plateaus by using the
idea proposed in \cite{hentschel:1983},

\begin{equation}
\left(\frac{S^{\prime}}{S}\right)^{D^{\prime}} + \left(\frac{S^{\prime\prime}}{S} \right)^{D^{\prime}} = 1.
\label{scaling_2}
\end{equation}
\noindent
where    $S^{\prime}$,    $S$,   and    $S^{\prime\prime}$    are   represented    in
Fig. \ref{meco_fig6_1}.

In case one has $K \cong 1$ ($K<1$), we do not have a complete devil's
staircase. In other words, winding numbers with denominators larger
than a given ${\widetilde{Q}}$ are cut off from the Farey Tree.
Using this information we can estimate the value of $K$ through the
largest denominator observed \cite{jensen:1984}

\begin{equation}
{\widetilde{Q}} \geq \frac{1}{1-K}. 
\label{cutoff}
\end{equation}

Finally, we would like to stress that while in Eq. (\ref{circle_map}) the
plateaus of the devil's staircase are positioned at winding numbers defined by
Eq. (\ref{winding_number}), in nature, devil's staircases have
plateaus positioned at some accessible measurement.

In the driven Rayleigh-B{\'e}nard experiment
\cite{stavans:1985,jensen:1985}, convection rolls appear in a small
brick-shaped cell filled with mercury, for a critical temperature difference
between the upper and lower plates. As one perturbs the cell by a constant
external magnetic field parallel to the axes of the rolls and by the introduction
of an AC electrical current sheet pulsating with a frequency $f_{e}$ and
amplitude $B$, a devil's staircase is found in the variable $f_i$, the main
frequency of the power spectra of the fluid velocity.  As one varies the
external frequency $f_{e}$, stable oscillations take place at a frequency
ratio $f_i/f_e$, for a given value of $f_e$.  In analogy to the devil's
staircase of Eq.  (\ref{circle_map}), $f_e$ should be thought as playing the
same role of $\Omega$ in Eq. (\ref{circle_map}), and the ratio $f_i/f_e$ as
playing the same role of the winding number $W$.

A devil's staircase can also be observed (see Ref. \cite{baptista:2004}) in
the amount of information $H$ (topological entropy) that an unstable chaotic
set has in terms of an interval of size $\epsilon$, used to create the set.
To generate the unstable chaotic set, we eliminate all possible trajectories
of a stable chaotic set that visits this interval $\epsilon$. In analogy
with the devil's staircase of the circle map, $\epsilon$ should be related to
$\Delta \Omega$, while $H$ to $W$.

The first proof of a complete devil's staircase in a physical model was given 
in Ref.  \cite{bak:1982}, in the one-dimensional ising model with convex
long-range antiferromagnetic interactions.

In Ref.
\cite{jin:1994}, it was found that a model for the El Ni{\~n}o, a
phenomenon that is the result of a tropical ocean-atmosphere interaction when
coupled nonlinearly with the Earth's annual cycle, could undergo a transition
to chaos through a series of frequency-locked steps.  The overlapping of these
resonances, which are the steps of the devil's staircase, leads to the chaotic
behavior.

\section{A devil's staircase in the Southern Brazilian Continental
  Shelf}\label{devil_in_data}

The main premise that guides the application of the devil's staircase model to
the Shelf is that the rules found in quantities related to the widths and
depths of the terraces obey the same rules found in a complete devil's
staircase for the frequency-locked intervals $\Delta \Omega$ and their
rational winding number, $W$. Thus, we assume that the terrace widths $l_n$
play the same role of the frequency-locked intervals $\Delta \Omega$, and the
terrace depths play the same role of the rational winding number $W$.

In order to interpret the Shelf as a devil's staircase, we have to show that
the terraces appear in positions which respect the Farey mediant, the rule
that describes the winding number "positions" of the many plateaus.  For that,
we verify whether the position of the terraces at $d_n$ can be associated to
hypothetical frequency ratios, denoted by $w_n=\frac{p_n}{q_n}$, which
respects the Farey mediant. In doing so, we want that the metric of three
adjacent terraces respects the Farey mediant. In addition, we also assume that
the larger terraces are the parents of the Farey Tree, while the smaller
terraces between two larger ones are the daughters. Thus, for each triple of
terrace, we want that

\begin{equation}
\frac{d_{n+2}+d_{n}}{d_{n+1}}= \frac{w_{n+2}+w_{n}}{w_{n+1}}.
\label{metric_equivalence}
\end{equation}
\noindent
One could have considered other ways to relate the depths and the frequency
ratios. The one chosen in Eq. (\ref{metric_equivalence}) is used in order to
account for the fact that while $d_n$ is negative $w_n$ is not.

From Eq. (\ref{metric_equivalence}) it becomes clear that for the further
analysis the depth of a particular terrace does not play a so important role
as the ratio between the depths of triple of terraces that contains this
particular terrace. These ratios may eliminate the influence of the local
morphology and the influence of the local sea level dynamics into the
formation of the Shelf. Therefore, the proposed quantity might be suitable for
an integrate analysis of the different Shelfs all over the world, specially
the ones affected by local geomorphological characteristics.

From the Farey mediant, we have a way to obtain the frequency ratios
associated to each terrace,

\begin{equation}
w_{n+1}=\frac{p_n + p_{n+2}}{q_n + q_{n+2}}.
\label{farey_rule}
\end{equation}
\noindent
Therefore,     combining     Eq.    (\ref{metric_equivalence})     and
Eq. (\ref{farey_rule}), we obtain

\begin{equation}
\frac{p_n  +  p_{n+2}}{q_n  +  q_{n+2}}=\frac{\frac{p_{n+2}}{q_{n+2}}+
\frac{p_{n}}{q_{n}}}{E},
\label{regra_terrace_1}
\end{equation}
\noindent
which results in

\begin{equation}
(p_n+p_{n+2})(E-1)q_{n}q_{n+2}p_{n}=p_{n}q_{n+2}^2 + q_n^2p_{n+2},
\label{regra_terrace_2}
\end{equation}
\noindent
where $E$ is defined by

\begin{equation}
\frac{d_{n+2}+d_{n}}{d_{n+1}}=E.
\label{regra_dados}
\end{equation}

\begin{center}
\begin{table}
\begin{tabular}{c|c|c} 
  $n$ & $p_n$  & $q_n$ \\ \hline
  1 & 1 & 8   \\
  2 & 2 & 17  \\
  3 & 1 & 9   \\
  4 & - & -   \\
  5 & - & -   \\
  6 & 1 & 17  \\
  7 & 2 & 35  \\
  8 & 1 & 18  \\ \hline
\end{tabular}
\caption{Integers associated with the $n$ considered terraces, with $n=1,\ldots,8$.}
\label{table2}
\end{table}
\end{center}

\begin{figure}
\centerline{\hbox{\psfig{file=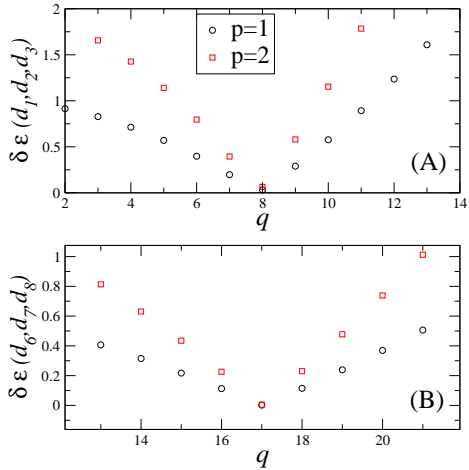,width=7.0cm}}} 
\caption{[Color online] (A) Values of $\delta \epsilon(d_1,d_2,d_3)$ and (B) 
$\delta \epsilon(d_6,d_7,d_8)$, for different values of $p$ and $q$. $\delta
\epsilon$ is the difference between the left and the right hand sides of
Eq. (\ref{regra_terrace_2}).}
\label{meco_fig20}
\end{figure}

We do not expect to have Eq.  (\ref{regra_terrace_2}) satisfied.  We only
require that the difference between the left and right hand sides of this
equation, regarded as $\delta \epsilon$, is the lowest possible, among all
possible values for $p_m$ and $q_m$ (with $m=n,n+2$), for a given $E$, with
the restriction that the considered largest terraces are related to largest
plateaus of Eq. (\ref{circle_map}), and thus $p_{m+2}$=$p_{m}$ and
$q_{m+2}$=$q_{m}+1$, and $\delta \epsilon \ll 1$.  Doing so, we find the
rationals associated with the terraces, which are shown in table
\ref{table2}. The minimal value of $\delta \epsilon$, denoted by 
$\min{[\delta \epsilon]}$, is $\min{[\delta
  \epsilon(d_1,d_2,d_3)]}$=0.032002, with $p_1/q_1$=$1/8$ for the terrace 1, and
$p_3/q_3$=$1/9$, for the terrace 3. We also find that $\min{[\delta
\epsilon(d_6,d_7,d_8)]}$=0.002344, 
with $p_6/q_6$=$1/17$, for the terrace 6, and $p_8/q_8$=$1/18$, for the terrace
8. These minimal values can be seen in Figs.  \ref{meco_fig20}(A-B), where we
show the values of $\delta \epsilon(d_1,d_2,d_3)$ [in (A)] and the values of
$\delta \epsilon(d_6,d_7,d_8)$ [in (B)], for different values of $p$ and $q$.
Using bigger values for $p$ has the only effect to increase the value of $\delta \epsilon$.  

We have not identified rationals that can be associated with the terraces $4$
and $5$, which means that for $p$ and $q$ within $p=[1,50]$ and $q=[1,400]$,
we find that $\delta
\epsilon > 1$. We have  assumed that they could be either a daughter or a parent.

From now on, when convenient, we will drop the index $n$ and represent each
terrace by the associated frequency ratio.  So, the terrace 1, for $n$=1, is
represented as the terrace with $w=1/8$.

\begin{figure}
\centerline{\hbox{\psfig{file=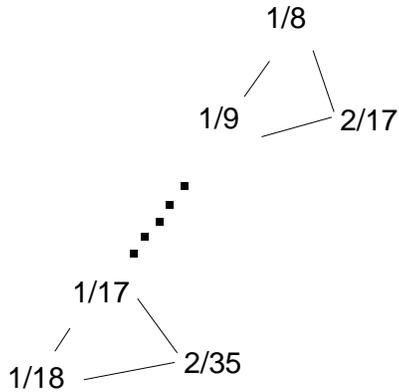,width=7.0cm}}} 
\caption{Farey Tree representing the frequency ratios associated with the
  major terraces.}
\label{meco_fig3}
\end{figure}

Table \ref{table2} can be represented in the form of the Farey Tree
as shown in Fig. \ref{meco_fig3}.  The branch of rationals in the
Farey Tree in the form $1/q$ belongs to the most stable branch, which
means that the observed terraces should have the largest widths.  
We believe that the other less important branches
of the complete devil's staircase present in the data were smoothed
out by the action of the waves and the flow streams throughout the
time, and at the present time cannot be observed.

Notice that as the time goes by, the frequency ratios are increasing
their absolute value, which means that if this tendency is preserved
in the future, we should expect to see larger terraces.

\begin{figure}
\centerline{\hbox{\psfig{file=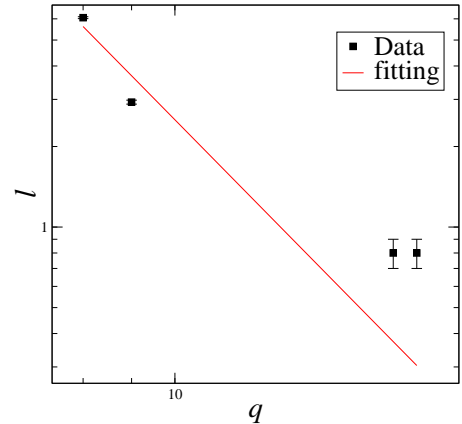,width=7.0cm}}} 
\caption{Scaling between the $1/q$-terrace widths and the value of
  $q$.}
\label{meco_fig5}
\end{figure}

In the following, we will try to recover in the experimental profile,
the universal scaling laws of Eqs.  (\ref{scaling_1}) and
(\ref{scaling_2}).  Regarding Eq.  (\ref{scaling_1}), we find that $l$
scales as $1/q^{-3.60}$, as shown in Fig. \ref{meco_fig5}, which is
the expected global universal scaling for a complete devil's
staircase. Regarding Eq.  (\ref{scaling_2}), and calculating
$S^{\prime}$, $S^{\prime\prime}$, and $S$ using the triple of terraces
with widths $l(w=1/8)$, $l(w=2/17)$, and $l(w=1/9)$, as represented in
Fig.
\ref{meco_fig6_1}, we find $D^{\prime}$=0.89. Using the triple of
terraces ($n$=3,$n$=4,$n$=5), we find that $D^{\prime}$=0.87. Both
results are very close from the universal fractal dimension $D_0 \cong
0.87$, found for a complete devil's staircase.

\section{Fitting the SBCS}\label{model}

\begin{figure}
\centerline{\hbox{\psfig{file=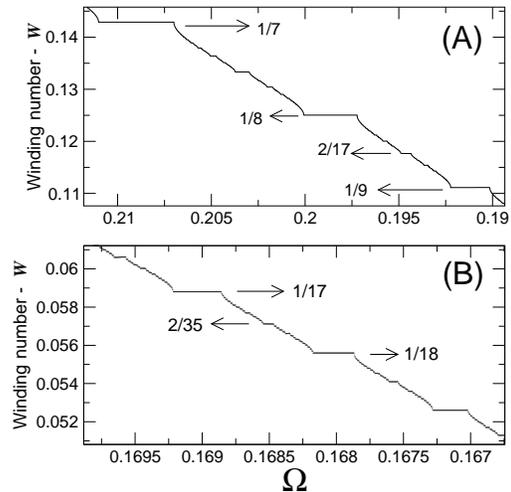,width=8.0cm}}} 
\caption{Magnifications of the small box of Fig. \ref{meco_fig6},
  showing the plateaus of the devil's staircase of Eq.
  (\ref{circle_map}) that appear for the same frequency ratios associated with the
  triple of terraces $w$=(1/8,2/17,1/9), in (A), and $w$=(1/17,2/35,1/18), in
  (B).}
\label{meco_fig7}
\end{figure}

Motivated by our previous results, we fit the observed Shelf as a
complete devil's staircase, using Eq. (\ref{circle_map}). Notice that
the only requirement for Eq. (\ref{circle_map}) to generate a complete
devil's staircase is that the function $g$ has a cubic inflection
point at the critical parameter $K=1$. Whether Eq. (\ref{circle_map})
is indeed an optimal modeling for the Shelf is beyond the scope of
the present study. We only chose this map because it is a well known
system and it captures most of the relevant characteristic a dynamical
systems needs to fulfill in order to create a devil's staircase.

We model the SBCS as a complete devil's staircase, but we rescale the
winding number $W$ into the observed terrace depth.  So, we transform
the complete devil's staircase of Fig. \ref{meco_fig7} as good as
possible into the profile of Fig.  \ref{meco_fig1}(B), by rescaling
the vertical axis of the staircase in Fig.  \ref{meco_fig7}.

\begin{figure}
\centerline{\hbox{\psfig{file=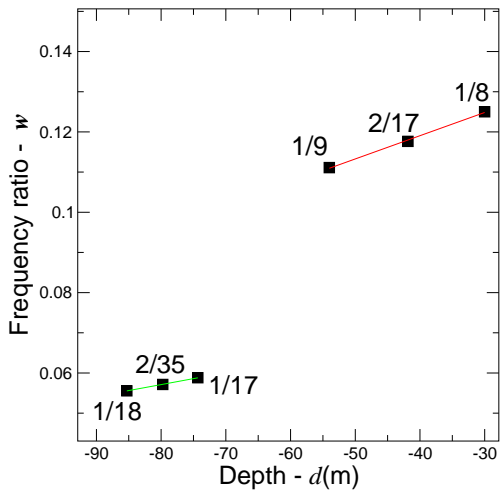,width=8.0cm}}} 
\caption{The function $F$, which is a linear relation 
between the frequency ratios associated with the terraces and their
depths for the triple of terraces $w$=(1/8,2/17,1/9) and 
$w$=(1/17,2/35,1/18).}
\label{meco_fig4}
\end{figure}

We do that by first obtaining the function $F$ (see Fig. \ref{meco_fig4})
whose application into the terrace depth $d(w)$
gives the frequency ratio $w_{n}=\frac{p_{n}}{q_{n}}$
associated with the terrace.

For the triple of terraces $w$=(1/8,2/17,1/9), we obtain 
\begin{equation}
F(d[km])=0.14219+0.00057853d[km], 
\label{function_F1}
\end{equation}
and for the triple of terraces $w$=(1/17,2/35,1/18), we obtain 
\begin{equation}
F(d[km])=0.080941+0.00029786d[km]. 
\label{function_F2}
\end{equation}
\noindent
Therefore, we assume that, locally, the frequency ratios are linearly
related to the depth of the terraces.

Then, we rescale the vertical axis of the staircases in
Figs. \ref{meco_fig7}(A-B) and calculate an equivalent depth, $d$, for 
the winding number $W$ by using 

\begin{equation}
d = F^{-1}(W),
\end{equation}

\begin{figure}
\centerline{\hbox{\psfig{file=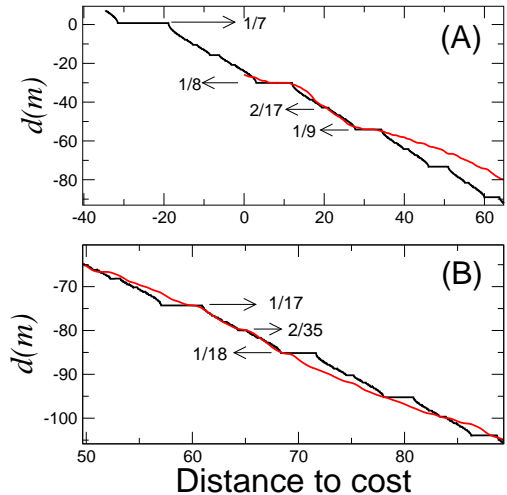,width=8.0cm}}} 
\caption{[Color online] (A) Rescaling of Fig. \ref{meco_fig7}(A) in black,
  showing that the devil's staircase fits well the terraces with
  $w$=(1/8,2/17,1/9) of the profile of Fig. \ref{meco_fig1}(B), in gray.  (B)
  Rescaling of Fig. \ref{meco_fig7}(B) in black, showing that the devil's
  staircase fits well the terraces with $w$=(1/17,2/35,1/18) of the profile of
  Fig. \ref{meco_fig1}(B), in gray.}
\label{meco_fig8}
\end{figure}

We also allow tiny adjustments in the axes for a best fitting.  The
result is shown in Fig.
\ref{meco_fig8}(A) for the triple of terraces $w$=(1/8,2/17,1/9)
and in Fig. \ref{meco_fig8}(B) for the triple of terraces
$w$=(1/17,2/35,1/18).  We see that locally, for a short time interval, we can
have a good agreement of the terrace widths and positions, with the rescaled
devil's staircase.  However, globally, the fitting in (A) does not do well, as
it is to be expected since the function $F$ is only locally well defined and it
changes depending on the depths of the terraces.  Notice however that this
short time interval is not so short since the time
interval correspondent to a triple of terraces is of the order of a few
hundred years.

The assumption made that $K=1$ is also supported from Eq.  (\ref{cutoff}).
Using this equation, we can obtain an estimation of the maximum value of $K$
from a terrace with a frequency ratio that has the largest denominator.  In our case,
we observed $w=2/35$.  Using ${\widetilde{Q}}$=35 in Eq.  (\ref{cutoff}), we obtain $K\leq
0.97$.

In Fig.  \ref{meco_fig8}(A), we see a 1/7 plateau positioned in the zero sea
level. That is the current level. Thus, the model predicts that nowadays we
should have a large terrace, which might imply in an average stabilization of
the sea level for a large period of time. However, this prediction might not
correspond to reality if the sea dynamics responsible for the creation of the
observed Continental Shelf suffered structurally modifications.

\section{Conclusions}\label{conclusao}

We have shown some experimental evidences that the Southern Brazilian
Continental Shelf (SBCS) has a structure similar to the devil's
staircase. That means that the terraces found in the bottom of the sea
are not randomly distributed but they occur following a dynamical
rule. This finding lead us to model the SBCS as a complete devil's
staircase, in which, between two real terraces, we suppose an infinite
number of virtual (smaller) ones. We do not find these later ones,
either because they have been washed out by the stream flow or simply
due to the fact that the time period in which the sea level dynamics
(SLD) stayed locked was not sufficient to create a terrace. By our
hypothesis, the SLD creates a terrace if it is a dynamics in which two
relevant frequencies are locked in a rational ratio.

This special phase-locked dynamics possesses a critical
characteristic: large changes in some parameter responsible for a
relevant natural frequency of the SLD might not destroy the
phase-locked regime, which might imply that the averaged sea level
would remain still.  On the other hand, small changes in the parameter
associated with an external forcing of the SLD could be catastrophic,
inducing a chaotic SLD, what would mean a turbulent averaged sea level
rising/regression.

In order to interpret the Shelf as a devil's staircase, we have shown
that the terraces appear in an organized way according to the Farey
mediant, the rule that describes the way plateaus appear in the
devil's staircase. That allow us to "name" each terrace depth, $d_n$,
by a rational number, $w_n$, regarded as the hypothetical frequency
ratio. Arguably, these ratios represent the ratio between real
frequencies that are present in the SLD. It is not the scope of the
present work to verify this hypothesis, however, one way to check if
the hypothetical frequency ratios are more than just a mathematical
artifact would be to check if the SLD has, nowadays, two relevant
frequencies in a ratio 1/7, as predicted.
 
The newly proposed approach to characterize the SBCS rely mainly on
the ratios between terraces widths and between terraces depths. While
single terrace widths and depths are strongly influenced by local
properties of the costal morphology and the local sea level
variations, the ratios between terrace widths and depths should be a
strong indication of the global sea level variations. Therefore, the
newly proposed approach has a general character and it seems to be
appropriated as a tool of analysis to other Continental Shelves around
the world.

Reminding that the local morphology of the studied area, the "Serra do
Mar" does not have a strong impact in the formation of the Shelf and
assuming that the local SLD is not directed involved in the formation
of the large terraces considered in our analyses, thus, our results
should reflect mainly the action of the global SLD.

If the characteristics observed locally in the S\~ao Paulo Bight
indeed reflect the effect of the global SLD, then the global SLD might
be a critical system. Hopefully, the environmental changes caused by
the modern men have not yet made any significant change in a relevant
parameter of this global system.

{\bf Acknowledgements:} MSB was partially supported by the ``Funda\c
c\~ao para a Ci\^encia e Tecnologia'' (FCT) and the Max-Planck
Institute f\"ur die Physik komplexer Systeme.


\begin{thebibliography}{99}
\bibitem{note3} The profiles used in this work were obtained by the
  use of polinomial spilines in a two dimensional grip of points 10km
  appart.  This also imposes limits in the indentification of terraces
  that have small widths.
\end{thebibliography}
\end{document}